# Symmetry-Protected Minimum of Four Conventional Weyl Points in Nonmagnetic Crystals

*Ze-Xin Xue[*], Ke-Xin Pang[*], Yun-Yun Bai, Yanfeng Ge, Yong Liu, and Yan Gao[†]*

State Key Laboratory of Metastable Materials Science and Technology & Hebei Key Laboratory of Microstructural Material Physics, School of Science, Yanshan University, Qinhuangdao 066004, China

Realizing nonmagnetic Weyl semimetals (WSMs) with the minimal number of conventional Weyl points (WPs) and a clean Fermi surface remains a central challenge. Here, combining symmetry analysis with first-principles calculations, we establish the definitive conditions under which a nonmagnetic crystal can host exactly four conventional ($C = \pm 1$) WPs, identifying 76 space groups in the spinless limit and 83 in the spinful case that allow this minimal configuration. Guided by this framework, we predict two previously unknown boron allotropes, P6-$B_{48}$ and TBIN-$B_{48}$, as ideal WSMs. Both exhibits precisely four isolated WPs near the Fermi level, with exceptionally clean electronic structures. Notably, the WPs in P6-$B_{48}$ are pinned to high-symmetry points, while those in TBIN-$B_{48}$ lie along high-symmetry lines, leading to distinct and experimentally accessible surface states, including single and double Fermi arcs. Our work provides a complete symmetry-based foundation and pristine material platforms for minimal Weyl physics.

---

[*]These authors contributed equally to this work.
[†]Corresponding authors: yangao9419@ysu.edu.cn

# 1. Introduction

The discovery of Weyl semimetals (WSMs) has revolutionized our understanding of gapless topological phases[1-4], providing a condensed matter realization of the relativistic Weyl fermions originally proposed in high-energy physics[5]. These materials are defined by the presence of Weyl points (WPs)—two nondegenerate band crossings that serve as monopoles of Berry curvature in momentum space[4, 6]. The existence of WPs is predicated on the breaking of either time-reversal symmetry ($\mathcal{T}$) and/or spatial inversion symmetry ($\mathcal{P}$)[4, 7]. The non-trivial topology of WSMs manifests in observable phenomena such as the chiral anomaly[8-10] and the formation of Fermi arcs on the surface boundaries, connecting the projections of WPs of opposite chirality[11-14].

Despite remarkable experimental progress, exemplified by the discovery of the TaAs family[11-14] and numerous subsequent realizations[15-17], a persistent challenge confronts the field. Most experimentally confirmed WSMs host a proliferation of WPs (e.g., 24 in TaAs[14, 18]), while related compounds exhibit comparable or greater multiplicities[4, 7]. This abundance, while topologically robust, introduces substantial complications. The electronic structure near the Fermi level ($E_F$) typically contains numerous trivial bands that coexist with the WPs, obscuring intrinsic topological phenomena[4, 7]. Moreover, many WPs reside at energies significantly displaced from the $E_F$, diminishing their influence on low-energy physics and complicating experimental detection of Fermi arcs[19]. These complexities severely hinder both fundamental investigations of Weyl fermion physics and potential device applications that might exploit topological protection. Consequently, the search for ideal WSMs, characterized by the minimal symmetry-allowed number of WPs and a clean electronic band structure, has become the "holy grail" of the field[20-23].

The Nielsen-Ninomiya no-go theorem[24-25] dictates that WPs must appear in pairs of opposite chirality to ensure a vanishing net topological charge in the Brillouin zone (BZ). In magnetic systems where $\mathcal{T}$ is broken, the minimal number of WPs is two, representing the "hydrogen atom" of magnetic WSMs[23, 26-29]. However, in non-magnetic systems preserving $\mathcal{T}$, the constraints are more severe. The $\mathcal{T}$-symmetry maps a WP at a non–time-reversal-invariant momentum $\mathbf{k}$ to a partner at -$\mathbf{k}$ with the same chirality. To satisfy the vanishing net charge condition[24-25], an additional pair of opposite chirality is required, establishing a fundamental minimum of four WPs[4, 22, 30]. Although several materials hosting four $|C| = 1$ WPs have been reported[22, 31-37], a complete theoretical framework for the symmetry-protected minimal four $|C| = 1$ WPs in non-magnetic materials has remained conspicuously absent. Moreover, the possibility that all four minimal WPs are symmetry-pinned to high-symmetry points has so far remained unexplored.

In this work, we address this fundamental gap by establishing a complete symmetry-based framework for realizing the symmetry-protected minimum of four $|C| = 1$ WPs in nonmagnetic crystals. By analyzing the encyclopedia of emergent particles and their corresponding little-group representations across all 230 space groups (SGs), we exhaustively identify 76 SGs in the spinless limit and 83 in the spinful case that can host this minimal Weyl configuration. Guided by this framework, we predict two novel boron allotropes, P6-B$_{48}$ (SG 168) and TBIN-B$_{48}$ (SG 114), as exemplary ideal WSMs. Both materials exhibit excellent mechanical, dynamical and thermal stability and host exactly four WPs in the entire BZ. Crucially, we demonstrate that the specific embedding of WPs, at high-symmetry points in P6-B$_{48}$ and on high-symmetry lines in TBIN-B$_{48}$, gives rise to unique surface spectroscopic signatures, including extended double Fermi arcs that mimic a single pair of $|C| = 2$ WPs and unconventional "$Z$"-shaped double Fermi arcs. Our work not only establishes a complete roadmap for the design of symmetry-protected minimal conventional WSMs in nonmagnetic systems but also provides ideal material platforms for exploring simplified Weyl physics.

## 2. Results

### 2.1 Symmetry Conditions for Minimal four conventional WP Configurations

We begin by establishing the symmetry requirements necessary to stabilize the symmetry-protected minimal number of conventional ($|C| = 1$) Weyl points (WPs) in nonmagnetic ($\mathcal{T}$-invariant) system. Let the four $|C| = 1$ WPs be located at momenta $\boldsymbol{k}_1$, $\boldsymbol{k}_2$, $\boldsymbol{k}_3$, and $\boldsymbol{k}_4$ ($\boldsymbol{k}_1 \neq \boldsymbol{k}_2 \neq \boldsymbol{k}_3 \neq \boldsymbol{k}_4$), with corresponding little groups denoted by $G_{\boldsymbol{k}_1}$, $G_{\boldsymbol{k}_2}$, $G_{\boldsymbol{k}_3}$, and $G_{\boldsymbol{k}_4}$. For a space group ($G$) to host exactly four $|C| = 1$ WPs, the following symmetry conditions must be satisfied:

(i) The space group $G$ must allow the existence of $|C| = 1$ WPs.

(ii) If $k_1$, $k_2$, $k_3$, and $k_4$ individually exist, then their little groups must be identical, $G = G_{k_1} = G_{k_2} = G_{k_3} = G_{k_4}$, in order to avoid generating additional symmetry-related WPs.

(iii) If two WPs, e.g., $k_1$ and $k_2$, individually exist, while $k_3$, and $k_4$ are related by a space-group operation $\mathcal{O}$ (such as proper rotations ($\mathcal{R}$), inversion ($I$), mirror ($\sigma$), time-reversal ($\mathcal{T}$) operation, or their combination), then the space group $G$ admits the coset decomposition $G = G_{k_1} = G_{k_2} = G_{k_3} \cup \mathcal{O} G_{k_3}$, with the little group of $k_4$ satisfying $G_{k_4} = \mathcal{O} G_{k_3} \mathcal{O}^{-1}$.

(iv) If the WPs form two symmetry-related pairs, such that $k_1$ and $k_2$ are related by an operation $\mathcal{O}_1$, while $k_3$ and $k_4$ are related by another operation $\mathcal{O}_2$, the space group decomposes as $G = G_{k_1} \cup \mathcal{O}_1 G_{k_1} = G_{k_3} \cup \mathcal{O}_2 G_{k_3}$, where the corresponding little groups satisfy $G_{k_2} = \mathcal{O}_1 G_{k_1} \mathcal{O}_1^{-1}$, $G_{k_4} = \mathcal{O}_2 G_{k_3} \mathcal{O}_2^{-1}$. Notably, the operations $\mathcal{O}_1$ and $\mathcal{O}_2$ must be either both improper or both proper, including their possible combinations with $\mathcal{T}$.

(v) If all four momenta are linked by symmetry operations $\mathcal{O}_3$, $\mathcal{O}_4$, and $\mathcal{O}_5$, the space group $G$ admits the decomposition $G = G_{k_1} \cup \mathcal{O}_3 G_{k_1} \cup \mathcal{O}_4 G_{k_1} \cup \mathcal{O}_5 G_{k_1}$, where $G_{k_2} = \mathcal{O}_3 G_{k_1} \mathcal{O}_3^{-1}$, $G_{k_3} = \mathcal{O}_4 G_{k_1} \mathcal{O}_4^{-1}$, $G_{k_4} = \mathcal{O}_5 G_{k_1} \mathcal{O}_5^{-1}$. Crucially, to satisfy the Nielsen-Ninomiya no-go theorem, exactly two of the operations $\mathcal{O}_3$, $\mathcal{O}_4$, and $\mathcal{O}_5$ must be improper (e.g., a mirror, inversion, or roto-inversion) to reverse the chirality of the WPs and ensure a vanishing net topological charge in the Brillouin zone.

Under these constraints, we performed a systematic screening of all 230 SGs using the encyclopedia of emergent particles[38], considering both the absence and the presence of spin–orbit coupling (SOC). The classification summarized in Table I and Table SIII of the Supplemental Material (SM) establishes the definitive search space for realizing symmetry-protected minimal conventional WSMs in nonmagnetic crystals. Specifically, we find that 76 SGs in the spinless limit and 83 SGs in the spinful case can host minimal WSMs featuring exactly four WPs.

## 2.2 Crystal Structure and Comprehensive Stability Analysis

Guided by the symmetry criteria above, we identify two previously unknown 3D boron allotropes, P6-$B_{48}$ and TBIN-$B_{48}$, that crystallize in space groups $P6$ (No. 168) and $P\bar{4}2_1c$ (No. 114), respectively, both belonging to our catalogue of 76 candidate SGs in the spinless system for the symmetry-protected minimal conventional WSMs.

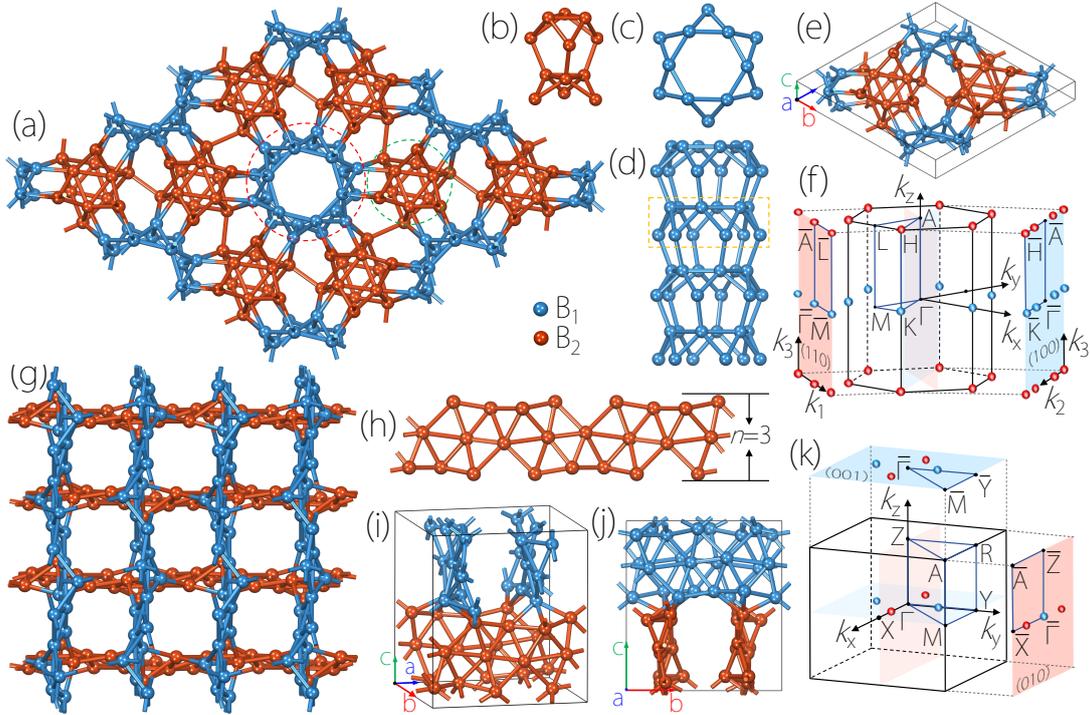

**Figure 1.** Crystal structures and Brillouin zones (BZ) of P6-$B_{48}$ and TBIN-$B_{48}$. (a) Top view

of a $2 \times 2 \times 2$ supercell of P6-B$_{48}$. (b) Side view of the green-dashed region in (a). (c) Top view of the distorted kagome lattice unit (yellow-dashed region in (d)). (d) Side view of the red-dashed region in (a), forming a nanotube-like structure. (e) Optimized primitive cell of P6-B$_{48}$. (f) Bulk BZ of P6-B$_{48}$ with four Weyl points (WPs) marked as red ($C = +1$) and blue ($C = -1$) spheres. Projections onto the (100) and (110) surfaces are shown. (g) Top view of a $2 \times 2 \times 2$ supercell of TBIN-B$_{48}$. (h) A triangular borophene nanoribbon with a width of three boron atoms, the building block of TBIN-B$_{48}$. (i) Perspective and (j) side views of the primitive cell of TBIN-B$_{48}$. (k) Bulk BZ of TBIN-B$_{48}$ with WPs and their projections onto the (001) and (010) surfaces.

**TABLE I.** Candidate space groups (SGs) that can host the symmetry-protected minimal configuration of four conventional Weyl points (WPs) with $|C| = 1$ (C-1 WPs) in nonmagnetic crystals, in both the spinless (without SOC) and spinful (with SOC) cases. PGs denote the point groups associated with the corresponding SGs, and SCs labels the corresponding symmetry conditions listed in Section A of the Results. Previously reported material realizations with symmetry-protected four-WP configurations, including Cu$_2$SnSe$_3$[33], MoTeSe[34], fco-C$_6$[35], and WTeS/WTeSe[36], as well as the currently proposed P6-B$_{48}$ and TBIN-B$_{48}$, are explicitly included. The specific momentum-space locations of the four WPs under their corresponding host SGs, together with the generators of these SGs, are summarized in Table SIII of the Supplemental Material.

| Four C-1 WPs | SGs (SCs) | PG | Example | SGs (SCs) | PG | Example |
|---|---|---|---|---|---|---|
| Without SOC | 3-5 (iv) | C$_2$ |  | 143-146 (iv) | C$_3$ |  |
|  | 16-24 (iv) | D$_2$ | fco-C$_6$ | 149-155 (iv) | D$_3$ |  |
|  | 35-37 (v), 44-46 (v) | C$_{2v}$ |  | 156 and 158 (v) | C$_{3v}$ |  |
|  | 75-80 (iv) | C$_4$ |  | 168-173 (iv) | C$_6$ | P6-B$_{48}$ |
|  | 81-82 (v) | S$_4$ |  | 174 (iv) | C$_{3h}$ |  |
|  | 89-98 (iv) | D$_4$ |  | 177-182 (iv) | D$_6$ |  |
|  | 111-122 (v) | D$_{2d}$ | TBIN-B$_{48}$ | 187-188 (iv) | D$_{3h}$ |  |
| With SOC | 1 (ii) | C$_1$ |  | 143-146 (iii,iv) | C$_3$ |  |
|  | 3-5 (ii,iii,iv) | C$_2$ |  | 149-155 (iii,iv) | D$_3$ |  |

| 8-9 (iv) | $C_s$ | WTeS/WTeSe; MoTeSe; Cu$_2$SnSe$_3$ | 156 and 158 (v) | $C_{3v}$ |
| --- | --- | --- | --- | --- |
| 16-24 (ii,iii,iv) | $D_2$ | | 168-173 (iii,iv) | $C_6$ |
| 35-37 (iv,v), 42-46 (v) | $C_{2v}$ | | 174 (iv) | $C_{3h}$ |
| 75-80 (ii,iii,iv) | $C_4$ | | 177-182 (iii,iv) | $D_6$ |
| 81-82 (iv,v) | $S_4$ | | 187-188 (iv) | $D_{3h}$ |
| 89-98 (ii,iii,iv) | $D_4$ | | 197 (iii) | $T$ |
| 111-122 (iv,v) | $D_{2d}$ | | 211 (iii) | $O$ |

The crystal structure of P6-B$_{48}$ is shown in Fig. 1(a). Its hexagonal unit cell [Fig. 1(e)] contains 48 boron atoms occupying eight distinct Wyckoff positions [see Table S1 in the SM]. The fully optimized lattice parameters are $a = b = 8.43$ Å and $c = 5.66$ Å. The structure can be visualized as an interlocking arrangement of two distinct structural motifs. The first is a quasi-one-dimensional nanotube formed by stacking distorted kagome lattices [Figs. 1(c) and 1(d)]. The second is composed of bilayer triangular boron units stacked with a 60° rotation along the $c$-axis [Fig. 1(b)]. The first BZ and the projection of WPs are shown in Fig. 1(f).

TBIN-B$_{48}$ (Triangular Borophene Interpenetrating Network) represents a distinct structural paradigm, constructed from interpenetrating networks of triangular borophene nanoribbons as shown in Fig. 1(g). This structure crystallizes in a primitive tetragonal lattice with a unit cell containing 48 boron atoms and optimized lattice constants $a = b = 7.15$ Å and $c = 7.85$ Å. The fundamental building block is a triangular borophene nanoribbon of width three atoms, illustrated in Fig. 1(h). These nanoribbons are arranged in a cross-stacked pattern, extending periodically along the $a$ and $b$ axes, and interconnected perpendicularly along the $c$ axis, as seen in the perspective and side views in Figs. 1(i) and 1(j). The tetragonal BZ, along with the WP locations and their projections onto the (001) and (010) surfaces, is presented in Fig. 1(k).

We comprehensively assessed the stability of both allotropes. The calculated elastic constants, listed in Table SII in the SM, satisfy the Born mechanical stability criteria[39] for their respective crystal systems, confirming their mechanical robustness. The phonon dispersion spectra, shown in Figs. 2(a) and 2(b), exhibit no imaginary frequencies throughout the BZ, proving their dynamical stability. Furthermore, AIMD simulations at 400 K for 8 ps [Figs. 2(c) and 2(d)] show that both structures maintain their integrity with only small fluctuations in total potential energy, indicating excellent thermal stability. These results strongly suggest that P6-$B_{48}$ and TBIN-$B_{48}$ are stable phases of boron and are viable candidates for experimental synthesis.

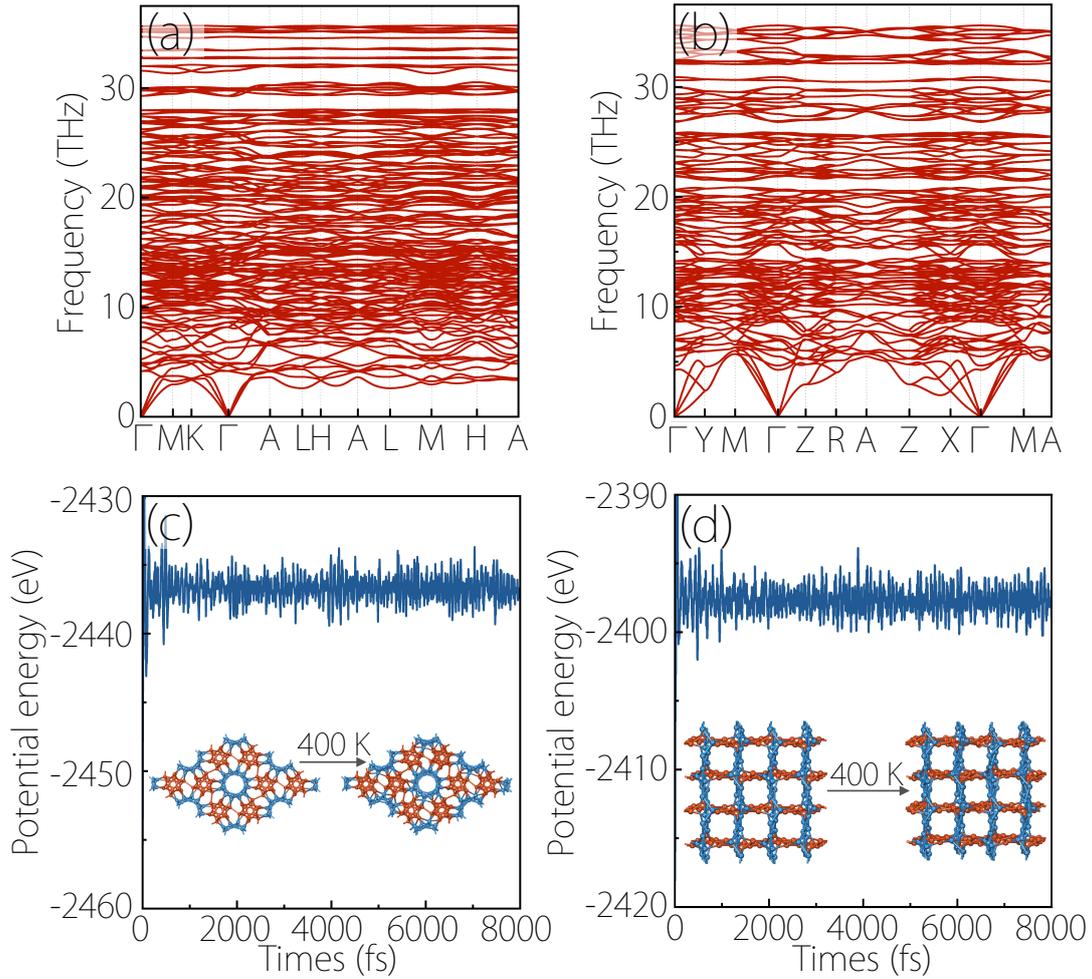

**Figure 2.** Stability analysis of P6-$B_{48}$ and TBIN-$B_{48}$. Phonon dispersion spectra of (a) P6-$B_{48}$ and (b) TBIN-$B_{48}$. No imaginary frequencies are present, confirming their dynamical stability. Total potential energy fluctuations during *ab initio* molecular dynamics (AIMD) simulations at 400 K for 8 ps for (c) P6-$B_{48}$ and (d) TBIN-$B_{48}$. Insets show the crystal structures before and after simulation, illustrating their thermal stability.

## 2.3 Topological Properties of P6-$B_{48}$: Minimal four WPs at High-Symmetry Points

We now turn to the electronic and topological properties of P6-$B_{48}$. Given the light mass of boron, spin-orbit coupling (SOC) effects are negligible, and the system can be treated as spinless. The calculated band structure along high-symmetry paths are shown in Fig. 3(a). The bands near $E_F$ are remarkably clean, governed solely by two linearly intersecting bands at the high-symmetry points H and K. These crossings form two distinct types of WPs, labeled $W_1$ (at H) and $W_2$ (at K). A global scan of the BZ confirms that these are the only four WPs present in the entire BZ. Due to $\mathcal{T}$ and crystal symmetries, there are two equivalent $W_1$ points at the H points and two equivalent $W_2$ points at the K points. Their locations and chiralities are summarized in Table II and depicted in Fig. 1(f).

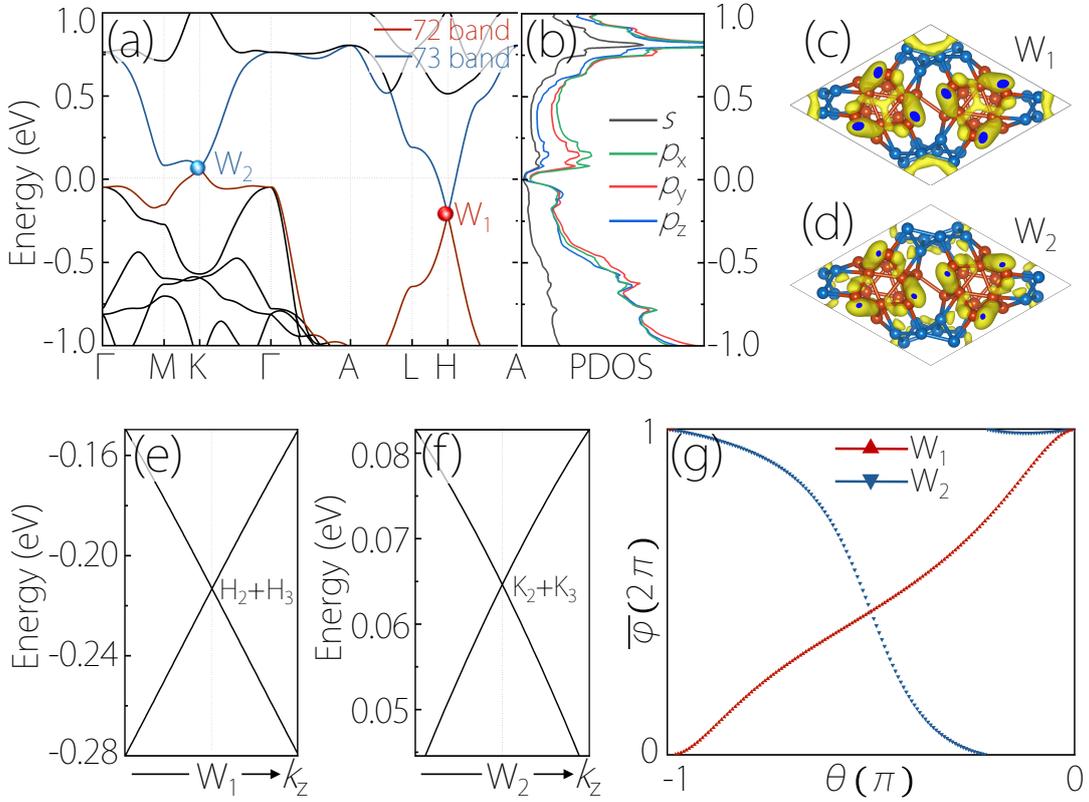

**Figure 3.** Electronic and topological properties of P6-$B_{48}$. (a) Bulk band structure along high-symmetry paths. (b) Projected density of states (PDOS). Partial charge density at the (c) $W_1$ and (d) $W_2$ points. Band dispersion along the $k_z$ direction through (e) $W_1$ and (f) $W_2$. Both Weyl points originate from distinct irreducible representations ($H_2 + H_3$ and $K_2 + K_3$). (g) Evolution of Wannier charge centers for spheres enclosing $W_1$ (red) and $W_2$ (blue), confirming their $C = +1$ and $C = -1$ chiralities.

The projected density of states (PDOS) in Fig. 3(b) reveals that the electronic states near $E_F$ are primarily derived from the boron $p$ orbitals. This is consistent with the charge density distributions at the WPs [Figs. 3(c) and 3(d)]. To confirm the $|C| = 1$ nature of these WPs, we examined their dispersion. The bands disperse linearly both in-plane [Fig. 3(a)] and out-of-plane [Figs. 3(e) and 3(f)], a hallmark of conventional $|C| = 1$ WPs. We explicitly calculated their topological charge by tracking the evolution of Wannier charge centers (WCCs) on enclosing spheres [Fig. 3(g)], which confirms that $W_1$ (at H) has a chirality of $C = +1$ and $W_2$ (at K) has a chirality of $C = -1$.

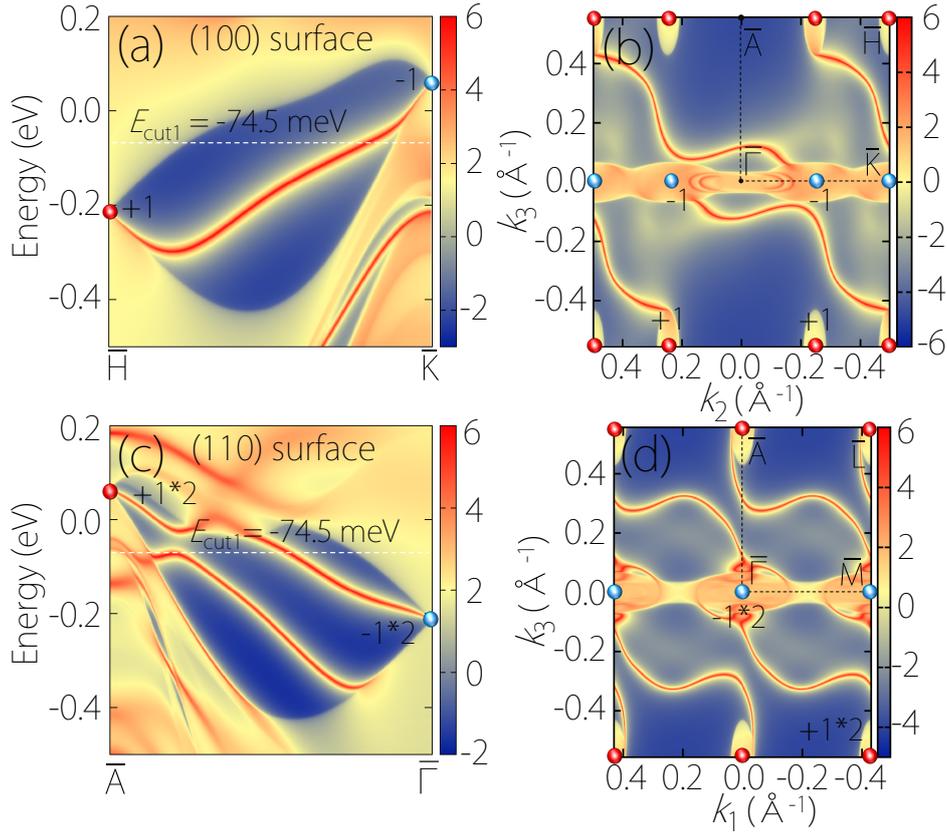

**Figure 4.** Surface electronic structures and Fermi-arc states of P6-B$_{48}$. (a) Surface band structure of the (100) surface calculated along the $\overline{H}$–$\overline{K}$ high-symmetry path. The dashed line marks the energy cut at $E_{\text{cut1}} = -74.5$ meV. (b) Constant-energy contour of the (100) surface at $E_{\text{cut1}}$, exhibiting two extremely long Fermi arcs. (c) Surface band structure of the

(110) surface along the $\overline{A}$–$\overline{\Gamma}$ direction. (d) Corresponding constant-energy contour for the (110) surface at $E_{\text{cut1}}$, revealing extended double Fermi arcs that connect the projected Weyl points.

The unique distribution of these four WPs at high-symmetry points leads to distinctive surface state signatures. We first consider the (100) surface, where the projections of $W_1$ and $W_2$ are well-separated [see Fig. 1(f)]. The calculated surface band structure [Fig. 4(a)] shows a clear topological surface state crossing $E_F$. The constant energy contour at $E - E_F = -74.5$ meV [Fig. 4(b)] reveals two extremely long Fermi arcs, each connecting a projected $W_1$-$W_2$ pair. These Fermi arcs extend across the full length of the surface BZ along the $k_3$ direction. The length and clarity of these Fermi arcs make them ideal targets for experimental observation via angle-resolved photoemission spectroscopy (ARPES).

A more remarkable feature appears on the (110) surface. Here, the two WPs of the same chirality (e.g., the two $W_1$ points with $C = +1$) are projected onto the same point in the surface BZ (the $\overline{A}$ point), while the two $W_2$ points ($C = -1$) project onto the $\overline{\Gamma}$ point [see Fig. 1(f)]. This projection results in an effective topological charge with $|C| = 2$ at $\overline{A}$ and $\overline{\Gamma}$. Consequently, the topological bulk-boundary correspondence requires two independent Fermi arcs connecting these projected charges. The calculated surface band structure for the (110) surface, displayed in Fig. 4(c), indeed shows two distinct surface state branches spanning the gap between $\overline{A}$ and $\overline{\Gamma}$ points. The constant energy contour in Fig. 4(d) clearly shows two parallel Fermi arcs, forming an extended double Fermi arc on the (110) surface. This feature, originating from the coalescence of projected WPs with identical chirality, closely resembles the surface states of a WSM hosting a single pair of $|C|=2$ WPs and serves as a distinctive spectroscopic fingerprint of P6-B$_{48}$.

## 2.4 Topological Properties of TBIN-B$_{48}$: Minimal four WPs on High-Symmetry Lines

We now investigate the tetragonal TBIN-B$_{48}$, which realizes minimal four WPs on high-symmetry lines rather than at high-symmetry points. Its bulk band structure [Fig. 5(a)] is also exceptionally clean near $E_F$, featuring two bands that cross precisely at the Fermi level. These crossings, W$_1$ and W$_2$, occur on the high-symmetry lines Γ-X and Γ-Y, respectively, realizing an ideal WSM state. A full BZ search confirms that there are only four WPs in total: one W$_1$ on Γ-X and its time-reversed partner, and one W$_2$ on Γ-Y and its time-reversed partner. All four WPs lie on the $k_z = 0$ plane. Their detailed information is summarized in Table II. The PDOS [Fig. 5(b)] and charge density plots [Figs. 5(c) and 5(d)] indicate that these states are also dominated by boron $p$ orbitals. The linear dispersion of these WPs is evident both in-plane [Fig. 5(a)] and out-of-plane [Figs. 5(e) and 5(f)]. We confirmed their topological charges using the Wilson loop method [Fig. 5(g)], finding that W$_1$ carries $C = +1$ and W$_2$ carries $C = -1$.

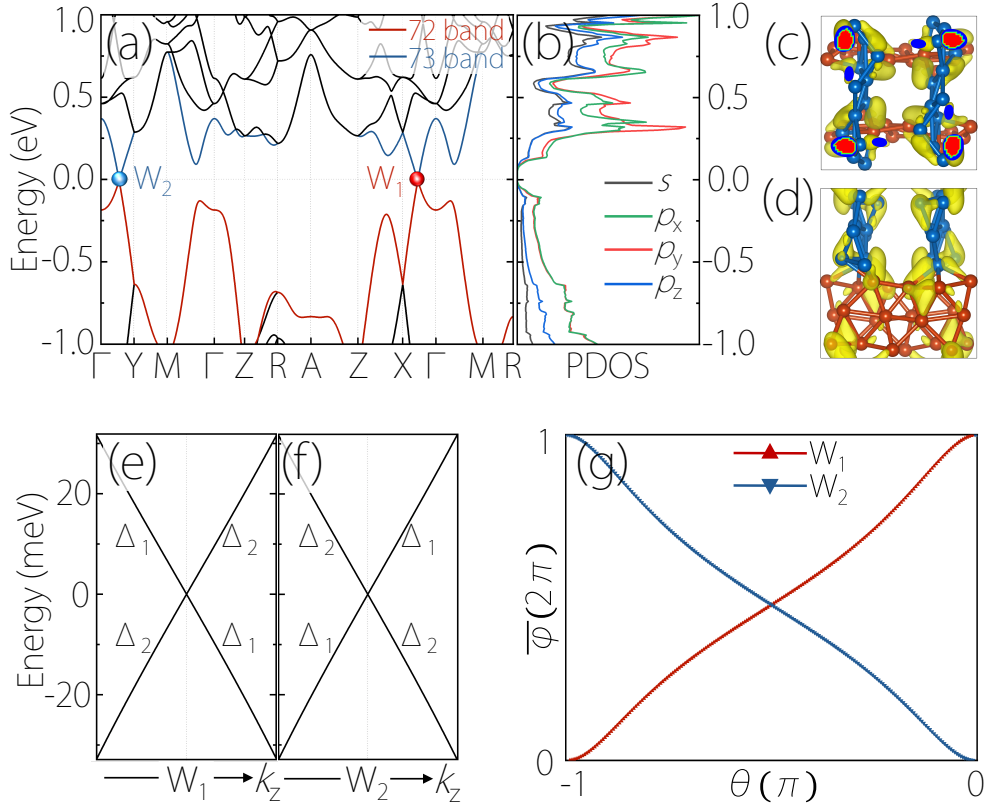

**Figure 5.** Electronic structure and topological characteristics of TBIN-$B_{48}$. (a) Bulk electronic band structure. (b) Projected density of states (PDOS). (c,d) Partial charge densities associated with the WPs $W_1$ and $W_2$, respectively. (e,f) Band dispersions along the $k_z$ direction passing through $W_1$ and $W_2$. Both WPs originate from distinct irreducible representations ($\Delta_1$ and $\Delta_2$). (g) Evolution of the Wannier charge centers for $W_1$ (red) and $W_2$ (blue), confirming their chiral charges of $C = +1$ and $C = -1$, respectively.

The surface states of TBIN-$B_{48}$ also exhibit unique features dictated by the WP locations. On the (001) surface, all four WPs project to distinct points [Fig. 1(k)]. The surface spectrum [Fig. 6(a)] shows clear topological states connecting the projections of oppositely charged WPs. The constant energy contour at $E_F$ [Fig. 6(b)] clearly illustrates two separate Fermi arcs connecting the two $W_1$-$W_2$ pairs.

**TABLE II.** Energies ($E$ (eV)), coordinates, topological charges, and multiplicities of the four Weyl points (WPs) in P6-$B_{48}$ and TBIN-$B_{48}$.

A more intriguing configuration arises on the (010) surface. Due to the crystal symmetry, the two $W_2$ points ($C = -1$) on the $\pm k_y$ axes project onto the same point, the surface BZ center ($\bar{\Gamma}$ point). Meanwhile, the two $W_1$ points ($C = +1$) on the $\pm k_x$ axes project to the $\bar{\Gamma} - \bar{X}$ line [Fig. 1(k)]. This creates a scenario where the $\bar{\Gamma}$ point has a projected charge of $|C| = 2$, while the two WPs on the $\bar{\Gamma} - \bar{X}$ line each have a charge of $|C| = 1$. The surface band structure [Fig. 6(c)] shows two states emerging from the $\bar{\Gamma}$ point. The constant energy contour [Fig. 6(d)] reveals a striking "Z"-shaped double Fermi arc pattern, with two arcs originating from $\bar{\Gamma}$ and connecting to the surface projections of the two $W_1$ points. This unique pattern is a direct consequence of the WP configuration and serves as an unambiguous

| Structures | WPs | $E$ (eV) | Coordinates ($k_1, k_2, k_3$) | Charges | Multiplicities |
|---|---|---|---|---|---|
| P6-B$_{48}$ | W$_1$ | 0.065 | ±(0.333, 0.333, 0.5) | +1 | 2 |
| | W$_2$ | −0.213 | ±(0.333, 0.333, 0) | −1 | 2 |
| TBIN-B$_{48}$ | W$_1$ | 0 | (±0.274, 0, 0) | +1 | 2 |
| | W$_2$ | 0 | (0, ±0.274, 0) | −1 | 2 |

experimental signature for TBIN-B$_{48}$.

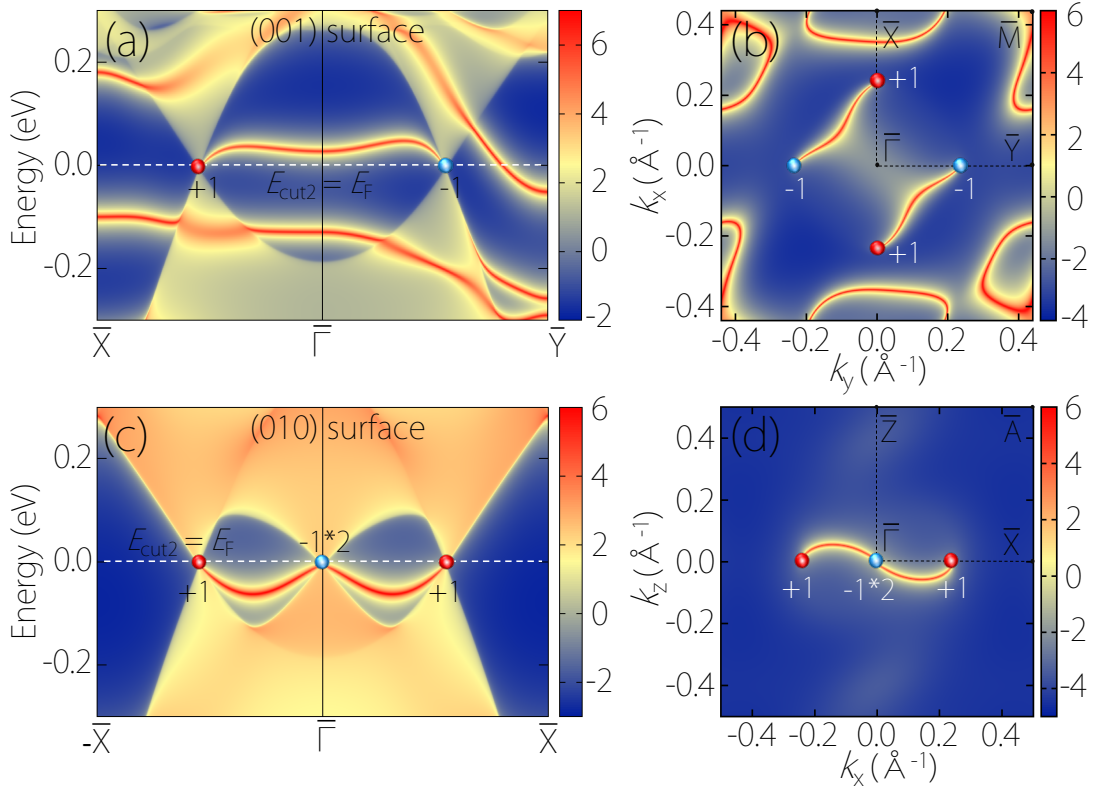

**Figure 6.** Surface states and Fermi arcs of TBIN-B$_{48}$. (a) Surface band structure of the (001) surface calculated along the projected $\bar{X}$–$\bar{\Gamma}$–$\bar{Y}$ path. The dashed line denotes the energy cut at $E_{\text{cut2}} = E_F$. (b) Constant-energy contour of the (001) surface at the Fermi level. (c) Surface band structure of the (010) surface along the projected -$\bar{X}$–$\bar{\Gamma}$–$\bar{X}$ direction. (d) Corresponding constant-energy contour of the (010) surface at $E_F$, revealing a unique "Z"-shaped double Fermi arc.

## 3. Discussion

The realization of ideal WSMs hinges on two central criteria: (i) achieving the smallest possible number of WPs consistent with the symmetry constraints of the system, and (ii) ensuring that these WPs are situated in a clean region of the Fermi surface near the Fermi level, free of trivial band interference. While magnetic WSMs can host a minimum of two WPs[23, 26-29], they are typically complicated by magnetic domains, magnetic scattering, and the challenge of stabilizing well-defined magnetic orders. By contrast, identifying minimal four-point WSMs in nonmagnetic systems[22, 31-35, 37] simplifies the symmetry landscape and enhances experimental accessibility, providing a clearer platform for observing intrinsic Weyl physics. Indeed, it is well known that the canonical example of a nonmagnetic minimal WSM, $TaIrTe_4$[22, 31-32], realizes exactly four WPs in its band structure, yet these WPs and their associated Fermi arcs lie entirely above the Fermi level, rendering them difficult to access using conventional spectroscopies.

For the second criteria, this condition is notoriously difficult to satisfy; many previously reported non-magnetic candidates with minimal WP counts suffer from WPs located far from the Fermi energy or heavily obscured by coexisting trivial bands, which short-circuit the topological transport channels (such as the chiral anomaly). In stark contrast, the boron allotropes identified here, P6-B48 and TBIN-$B_{48}$, exemplify the ideal limit: they not only strictly adhere to the symmetry-enforced minimum of four WPs but also possess an exceptionally clean band topology. In both materials, the WPs are located at or in the immediate vicinity of the Fermi level, entirely isolated from interfering trivial states. This intrinsic electronic cleanliness greatly facilitates the disentanglement of Weyl physics using ARPES and transport measurements, establishing these materials as ideal platforms for exploring nonmagnetic conventional Weyl fermions in their simplest form. Crucially, the identification of 76 SGs in the spinless limit and 83 in the spinful case provides a definitive search dictionary for future materials discovery, enabling a transition from serendipitous findings to symmetry-guided design.

## 4. Conclusions

In summary, we establish a complete symmetry-based framework for realizing the minimal number of conventional ($|C| = 1$) WPs in nonmagnetic systems and identify 76 SGs in the spinless limit and 83 in the spinful case that are compatible with this constraint. Guided by this theory, we predict two stable boron allotropes, P6-$B_{48}$ and TBIN-$B_{48}$, which host exactly four WPs within exceptionally clean electronic structures near the Fermi level. The discovery of WPs pinned to high-symmetry points and lines, together with their distinctive surface Fermi-arc signatures, provides clear experimental targets and advances the search for ideal WSMs. Consequently, our work not only expands the landscape of topological boron materials but also provides a systematic roadmap for exploring symmetry-protected minimal configurations of conventional WPs in nonmagnetic systems.

## 5. Methods

First-principles calculations were carried out within density-functional theory (DFT) as implemented in the VASP package[40] using projector augmented wave method[41]. The exchange-correlation functional was treated within the generalized gradient approximation (GGA) of the Perdew–Burke–Ernzerhof (PBE) type[42]. A plane-wave kinetic energy cutoff of 520 eV was employed. Brillouin-zone integrations were performed using Monkhorst–Pack $k$-point meshes[43] of $6 \times 6 \times 8$ for P6-$B_{48}$ and $6 \times 6 \times 6$ for TBIN-$B_{48}$. The convergence criteria for total energy and atomic forces were set to $10^{-6}$ eV and $10^{-3}$ eV/Å, respectively. Phonon dispersion spectra of P6-B$_{48}$ and TBIN-B$_{48}$ were computed using the finite-displacement method as implemented in the PHONOPY package[44] on $2 \times 2 \times 2$ supercells. *Ab initio* molecular dynamics (AIMD) simulations were conducted in the canonical (NVT) ensemble using a Nosé-Hoover thermostat[45]. Elastic constants ($C_{ij}$) obtained from energy-strain relations[46] and were found to satisfy the Born mechanical stability criteria. The topological properties of the P6-B$_{48}$ and TBIN-B$_{48}$ structures were analyzed using the Wannier90[47] and WannierTools Packages[48].

## Supporting Information

Supporting Information is available from the Wiley Online Library or from the author.

## Acknowledgements

We wish to thank Weikang Wu for helpful discussions. This work was supported by the National Natural Science Foundation of China (Grants No. 12304202), Hebei Natural Science Foundation (Grant No. A2023203007), Science Research Project of Hebei Education Department (Grant No. BJK2024085), and Cultivation Project for Basic Research and Innovation of Yanshan University (No. 2022LGZD001).

## Conflict of Interest

The authors declare no conflict of interest.

## Data Availability Statement

The data that support the findings of this study are available from the corresponding author upon reasonable request.